\PassOptionsToPackage{table,dvipsnames}{xcolor}  

\documentclass[manuscript]{acmart}

\AtBeginDocument{%
  }

\copyrightyear{2025}
\acmYear{2025}
\setcopyright{rightsretained}
\acmConference[CUI '25]{Proceedings of the 7th ACM Conference on Conversational User Interfaces}{July 8--10, 2025}{Waterloo, ON, Canada}
\acmBooktitle{Proceedings of the 7th ACM Conference on Conversational User Interfaces (CUI '25), July 8--10, 2025, Waterloo, ON, Canada}
\acmDOI{10.1145/3719160.3737636}
\acmISBN{979-8-4007-1527-3/2025/07}

\definecolor{softpink}{HTML}{F4A8A8}

\newcommand{\pinkcircled}[1]{%
  \begingroup
    \setlength{\fboxsep}{1pt}%
    \colorbox{softpink}{\textcolor{white}{\textbf{#1}}}%
  \endgroup
}

\usepackage{graphicx}
\usepackage{caption}
\usepackage{subcaption}
\usepackage{enumitem}
\usepackage{float}
\usepackage[table]{xcolor}
\usepackage{amsmath}

\acmSubmissionID{7082}

\begin{document}





\title{Feedstack: Augmenting Conversational Interfaces with Structured Layers for Feedback and Exploration}

\title{Feedstack: Layering Structured Representations over Unstructured Feedback to Support Exploration and Scaffold Human–AI Conversation}

\title{Feedstack: Layering Structured Representations over Unstructured Feedback to Scaffold Human–AI Conversation}




\author{Hannah Vy Nguyen}
\orcid{}
\affiliation{%
  \institution{Temple University}
  \city{Philadelphia}
  \state{PA}
  \country{USA}
}
\email{hannahnguyen4@sigchi.org}
\orcid{}

\author{Yu-Chun Grace Yen}
\affiliation{%
  \institution{University of San Diego}
  \city{La Jolla}
  \state{CA}
  \country{USA}
}
\email{yyen@ucsd.edu}
\orcid{}

\author{Omar Shakir}
\affiliation{%
  \institution{Temple University}
  \city{Philadelphia}
  \state{PA}
  \country{USA}
}
\email{omar.shakir@temple.edu}
\orcid{}

\author{Hang Huynh}
\affiliation{%
  \institution{Temple University}
  \city{Philadelphia}
  \state{PA}
  \country{USA}
}
\email{hang.huynh@temple.edu}
\orcid{}

\author{Sebastian Gutierrez}
\affiliation{%
  \institution{Temple University}
  \city{Philadelphia}
  \state{PA}
  \country{USA}
}
\email{guts@temple.edu}
\orcid{}

\author{June A. Smith}
\affiliation{%
  \institution{Berea College}
  \city{Berea}
  \state{KY}
  \country{USA}
}
\email{smithj7@berea.edu}
\orcid{} 

\author{Sheila Jimenez}
\affiliation{%
  \institution{Temple University}
  \city{Philadelphia}
  \state{PA}
  \country{USA}
}
\email{sheila.jimenez@temple.edu}
\orcid{}

\author{Salma Abdelgelil}
\affiliation{%
  \institution{Temple University}
  \city{Philadelphia}
  \state{PA}
  \country{USA}
}
\email{salmaabdelgelil@temple.edu}
\orcid{}

\author{Stephen MacNeil}
\affiliation{%
  \institution{Temple University}
  \city{Philadelphia}
  \state{PA}
  \country{USA}
}
\email{stephen.macneil@temple.edu}
\orcid{0000-0003-2781-6619}
\renewcommand{\shortauthors}{Hannah Vy Nguyen et al.}

\begin{abstract}

Many conversational user interfaces facilitate linear conversations with turn-based dialogue, similar to face-to-face conversations between people. However, digital conversations can afford more than simple back-and-forth; they can be layered with interaction techniques and structured representations that scaffold exploration, reflection, and shared understanding between users and AI systems.
We introduce Feedstack, a speculative interface that augments feedback conversations with layered affordances for organizing, navigating, and externalizing feedback. 
These layered structures serve as a shared representation of the conversation that can surface user intent and reveal underlying design principles. 
This work represents an early exploration of this vision using a research-through-design approach. We describe system features and design rationale, and present insights from two formative (n=8, n=8) studies 
to examine how novice designers engage with these layered supports. Rather than presenting a conclusive evaluation, we reflect on Feedstack as a design probe that opens up new directions for conversational feedback systems.

\end{abstract}



\maketitle

 \section {Introduction}
 
The adoption of conversational user interfaces (CUIs) in education is rapidly growing, especially as students increasingly rely on these tools as a help resource~\cite{hou2024effects, hou_evolvingusage_2025} and for feedback ~\cite{zaphiris_envisioned_2021, xiao_studyofefllearnersuseofchatgpt_2023}. This popularity reflects their accessibility and ability to provide immediate responses, often mimicking the back-and-forth flow of human dialogue. However, they are primarily designed to support linear conversations ~\cite{jain_evaluating_2018}, with few opportunities for divergence, reflection, or revisiting earlier points. As a result, conversations often remain confined to a single, linear path.

In contrast, a few information exploration tools are emerging that prioritize structure over conversation. These diagrammatic systems, such as CausalMapper or Sensecape ~\cite{huang2023causalmapper, suh2023sensecape}, offer affordances for exploring connections between ideas, but they sacrifice the fluid, iterative nature of dialogue. This demonstrates a current gap in design approaches, which we explore as the following research question: 

\begin{itemize}
    \item[\textbf{RQ:}] How might we adapt conversational user interfaces to better support exploration and reflection while maintaining the unstructured nature of dialogue?   
\end{itemize}

Some chatbots currently support exploration by mentioning related topics or offering follow-up questions; however, the burden is still placed on the user to notice connections and pursue them. This limitation is especially problematic for non-expert users, who may struggle to articulate what they need or to understand the rationale behind the feedback they receive~\cite{foong2017novice}. Without conversational cues or scaffolds that encourage reflection and exploration, users may accept feedback passively, without internalizing it, understanding it, or relating it to their broader goals.

In this paper, we investigate the potential to layer structured representations to support exploration and reflection. These `shared representations also serve to scaffold human-AI interaction, where implicit aspects of the conversation become explicit. To investigate this potential, we introduce \textbf{Feedstack}, a multimodal conversational user interface that actively supports exploration and reflection, while also maintaining the natural flow of the conversation. This design probe~\cite{boehner2012probes} was developed based on design insights from a formative study. Through a second user study, we identify additional opportunities for refinement, which have been integrated into a web-based prototype. This research-through-design approach~\cite{zimmerman2007research} is intended to advance our vision for chatbots that are anchored to shared representations that externalize information that is often implicit within the conversation or slightly outside the conversational context.

\section{Related Work} 



\subsection{The Nature of Feedback Conversations} 

Feedback is an essential component of effective learning, especially in domains like design, where iterative improvement is a critical skill to develop and students learn through critique from experts and peers~\cite{yen_listen_2017}. 
However, students often struggle to interpret and apply unstructured feedback.
Students also face challenges in knowing which parts of the feedback to prioritize~\cite{blair_what_2013} and how specific instances of the feedback relate to the underlying design principles~\cite{hattie_power_2007}. 
Previous research has shown that structured feedback (i.e., feedback that is clearly organized and tied to specific learning objectives) improves comprehension and application, which has led to numerous intelligent creativity support tools that focus on structuring unstructured feedback~\cite{fraser2017critiquekit, xu_voyant_2014}.

Feedback conversations are highly interactive and can meander into unexpected topics and critiques at any point during the conversation. 
For example, when an expert provides feedback, a student might respond with a clarifying question, especially when the feedback introduces unfamiliar topics or concepts. Due to the `curse of expertise'~\cite{hinds1999curse}, these disconnects between the ideas being expressed by experts and how they are interpreted by novices are common.

One approach to help students interpret feedback is to connect feedback instances to high-level design principles \cite{hinds_bothered_2001}. Prior work suggests that this can help students understand the broader context and underlying justifications behind the feedback~\cite{chan_formulating_2024}. Effective feedback also encourages students to reflect on how their work fits into `the big picture' by helping learners see their work from various perspectives and see connections in the feedback.  Unfortunately, in unstructured feedback, these connections are often implicit. 
We explore how mapping and externalizing high-level design principles in feedback can enhance student learning, particularly in chatbot conversations, by helping students better connect feedback to the broader design context and improve their decision-making.

\subsection{Structuring Unstructured Design Feedback}


Chatbots are increasingly used in educational settings to support teachers and students~\cite{roca_impactofachatbotworking_2024, sonderegger2022chatbot, liu_aiintelligencechatbot_2022}. Many students now turning to general purpose chatbots, such as ChatGPT, to answer their questions, explain content, and guide their learning~\cite{hou2024effects, hou_evolvingusage_2025}. These interfaces are familiar and largely replicate some of the natural language interactions they might have with peers and instructors. Recently, the performance of these chatbots has improved dramatically, requiring fewer explicit rules and less fine-tuning. The release of multimodal models now enables users to incorporate visual imagery in their chatbot conversations;for example, visual question answering~\cite{lee_realizingvisualquestion_2024}. 

However, research on intelligent feedback systems has consistently emphasized the importance of structuring feedback to improve learning and sense-making~\cite{xu_voyant_2014, luther2015structuring, yuan2016almost, krause2017critique, andolina2017crowdboard, wauck2017class, luther2014crowdcrit}. For example, Sensecape, CausalMapper, and Graphologue address the challenge of information exploration by providing multi-level abstractions and interactive diagrams \cite{suh_sensecape_2023, huang2023causalmapper, jiang_graphologue_2023}. Similarly, Decipher and Voyant provide feedback scaffolding by structuring feedback based on topics and intentions to help learners act on feedback more effectively \cite{yen2020decipher, xu_voyant_2014}.

We have identified a research gap in designing chatbots that support fluid feedback conversations, in a way that enables users to layer structured insights on top of unstructured conversations. 
Specifically, there is a need for CUIs that maintain the natural flow of feedback while also encouraging students to explore feedback topics concurrently, rather than in a strictly linear manner. 



\section {Formative Study}

To explore how novice designers engage with traditional CUIs in a design feedback context, we conducted a formative study. Our study investigated how standard chatbot interfaces (i.e. ChatGPT) facilitates or fails to facilitate feedback interactions for non-expert users. Participants were tasked with engaging in a feedback conversation with ChatGPT to stimulate a design critique session. 
Designers increasingly interact with chat interfaces that are integrated within their design tools (e.g.: Figma); however, these chat interfaces are designed to support design activity directly, and not to provide feedback. Designers often still rely on external feedback sources such as a peer, advisor, or ChatGPT.

We recruited 8 university-level students from various institutions, including 6 undergraduates and 2 graduates from diverse majors, such as Chemistry, Education, Computer Science, and Psychology. Participants were recruited via university mailing lists and communication channels, including Slack. Each participant received \$20 USD compensation in the form of a digital gift card for their time. We did not require any prior experience with formal feedback systems or ChatGPT. Participants were provided with a fictional website mockup and instructed to imagine themselves as beginner designers revising their interface based on expert feedback. Participants could freely interact with ChatGPT and the sessions lasted 7-10 minutes. The team qualitatively coded the conversations to understand the patterns and dynamics of how the participants interacted with the ChatGPT interface. The study was approved by IRB.

Our main findings suggest that users use ChatGPT for requesting examples and asking clarifying questions. They sometimes had trouble leading the conversation and other times changed topics to guide their exploration of design topics. Based on these findings and the existing literature, we devised the following design goals:

\begin{quote}
\textbf{D1. Preserve the Open-Ended Nature of Feedback} Retain the unstructured, conversational quality of feedback discussions to support ambiguity, nuance, and interpretation.  
\end{quote}

\begin{quote}
\textbf{D2. Support Exploration} Enable users to connect individual pieces of feedback to broader design principles. Encourage discovery of relevant but undiscussed principles through prompts, suggestions, or visualizations that surface thematic gaps. 
\end{quote}

\begin{quote}
\textbf{D3. Promote Reflection} Help users track what has been discussed (and what has not) across the conversation. Surface patterns of topics or absences to promote reflection.
\end{quote}

\section{{Feedstack Design}}

\begin{figure}
    \centering
    \includegraphics[width=1\linewidth]{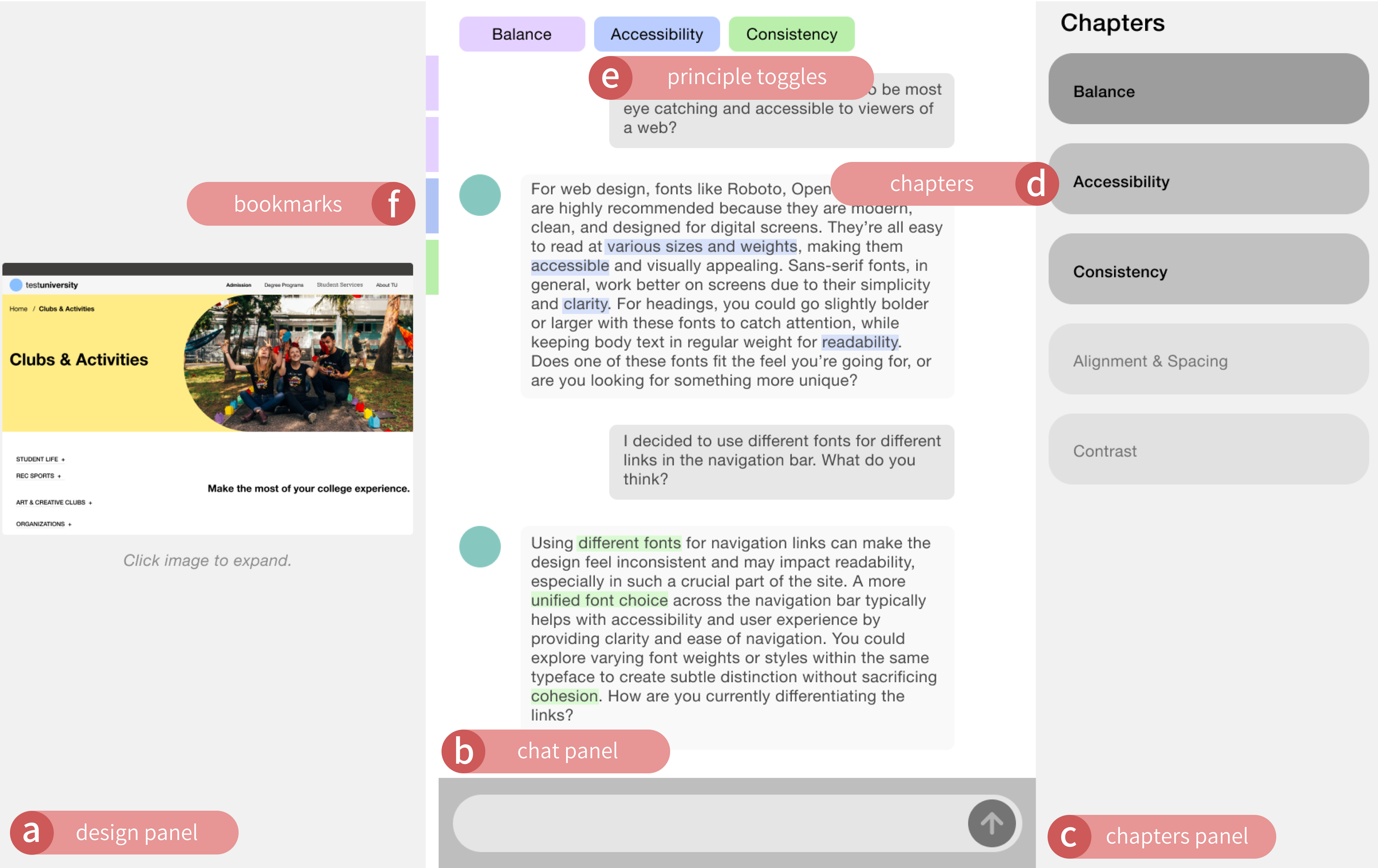}
    \caption{The design panel \protect\pinkcircled{A} displays the user's design artifact. The chat panel \protect\pinkcircled{B} contains the chat where the user can interact with the LLM-powered design expert to receive and explore feedback. The chapters panel \protect\pinkcircled{C} contains the interactive chapters \protect\pinkcircled{D}. On the top of the chat panel is the principle toggles \protect\pinkcircled{E} and on the left of the chat panel is the bookmarks \protect\pinkcircled{F}.}
    \label{fig:system}
    \Description{Screenshot of the Feedstack interface with three main panels. The left panel (A) shows the user’s design artifact. The center panel (B) displays the chat interface where users interact with a design expert chatbot. The right panel (C) shows interactive chapters (D), each representing a visual design principle. At the top of the chat panel are principle toggles (E), and on the left side of the chat panel are bookmarks (F) indicating when each principle was discussed.}
\end{figure}

We designed an interactive system called \textbf{Feedstack}, an interface that scaffolds real-time feedback conversations between the user and the LLM through interactions and affordances outlined below. The name relates to the idea that structured feedback is \textit{`stacked'} on top of an unstructured feedback conversation, kind of like how an integrated development environment (IDE) provides affordances around the code. The chatbot interface is in the center (See \pinkcircled{B} in Figure~\ref{fig:system}) with peripheral panels on the left and right. This conversation revolves around a visual design, which users can upload and view through the conversation (See \pinkcircled{A} in Figure~\ref{fig:system}).

\subsection{Bookmarks}

The Bookmarks feature (See \pinkcircled{F} in Figure~\ref{fig:system}) visually marks points along the discussion’s scrub bar where specific design principles were discussed. By showing these segments, it allows users to quickly locate and revisit moments when particular principles were discussed, similar to how `Find' features in PDF viewers highlight all occurrences of a keyword in a scroll bar. 

\subsubsection{Design Rationale}

This design supports \textit{reflection}~\cite{schon2017reflective} by allowing novice designers to easily revisit key parts of the conversation, enabling focused review and deeper thinking about each principle. It also reveals the extent to which each principle was discussed, highlighting potential over- or under-emphasis. The design also aligns with \textit{variation theory}~\cite{gu2004teaching, margulieux2021wrong} by helping users quickly navigate to multiple examples of how a given principle was applied. Juxtaposing these instances allows learners to notice subtle differences in interpretation and application across contexts.

\subsection{Chapters}

As design principles are identified in the conversation, they appear on the side panel as Chapters (See \pinkcircled{D}) to help make these implicitly discussed design principles explicit to users. Chapters are generated in real time using a large language model to analyze the conversation. In our initial prototype, the chapters were represented as five pre-defined accordions, each representing a core principle derived from established visual design principles~\cite{lidwell2020universal}. Chapters also include the following features:   

\begin{itemize}
    \item \textit{Learning Materials:} Each expanded chapter is organized into three sections: 1) \textit{Principle Definition:} A clear explanation of the principle, 2) \textit{Relation to Your Design}: Insights about how the principle applies to the user’s design, 3) \textit{Key Terms}: A glossary of important terminology related to the principle. These materials are generated using a large language model when the design principle is first identified in the conversation.
    \item \textit{Collapsibility:} To minimize cognitive load, all chapters are initially presented in a collapsed state. This design reduces the visible information, helping users focus on specific topics without feeling overwhelmed. Users can open or close the accordions at any time. 
    \item \textit{Opacity} As users interact with each principle during feedback conversations, the opacity of the corresponding chapter gradually increases. This increase in opacity reflects the frequency with which the principle is referenced, signaling to users and the chatbot which topics are of interest and which are underexplored. 
\end{itemize}

\subsubsection{Design Rationale}

The Chapters were designed as a shared representation between the AI and user. In mixed-initiative systems, \textit{shared representations}~\cite{heer2019agency} can provide common ground between the agent and the user, facilitating mutual understanding. In the context of feedback conversations, the Chapters externalize the design principles that often remain tacit or implicit within natural language exchanges. This externalization helps users zoom out from the back-and-forth flow of conversation and see the larger structure and themes that are unfolding.

\subsection{Highlights} 

Key terms related design principles are highlighted to focus the user's attention on important areas of feedback. As new design principles emerge in the discussion, they are  listed at the top of the chat conversation (see \pinkcircled{E} in Figure~\ref{fig:system}). By clicking on them, users can turn the highlights on, or turn them off to reduce their cognitive load.  

\subsubsection{Design Rationale}

Drawing on dual-coding theory \cite{clark_dualcoding_1991}, the highlights augment text-based feedback with visual cues to help users better understand which design principles are being emphasized during the conversation.

\begin{figure}
    \centering
    \includegraphics[width=1\linewidth]{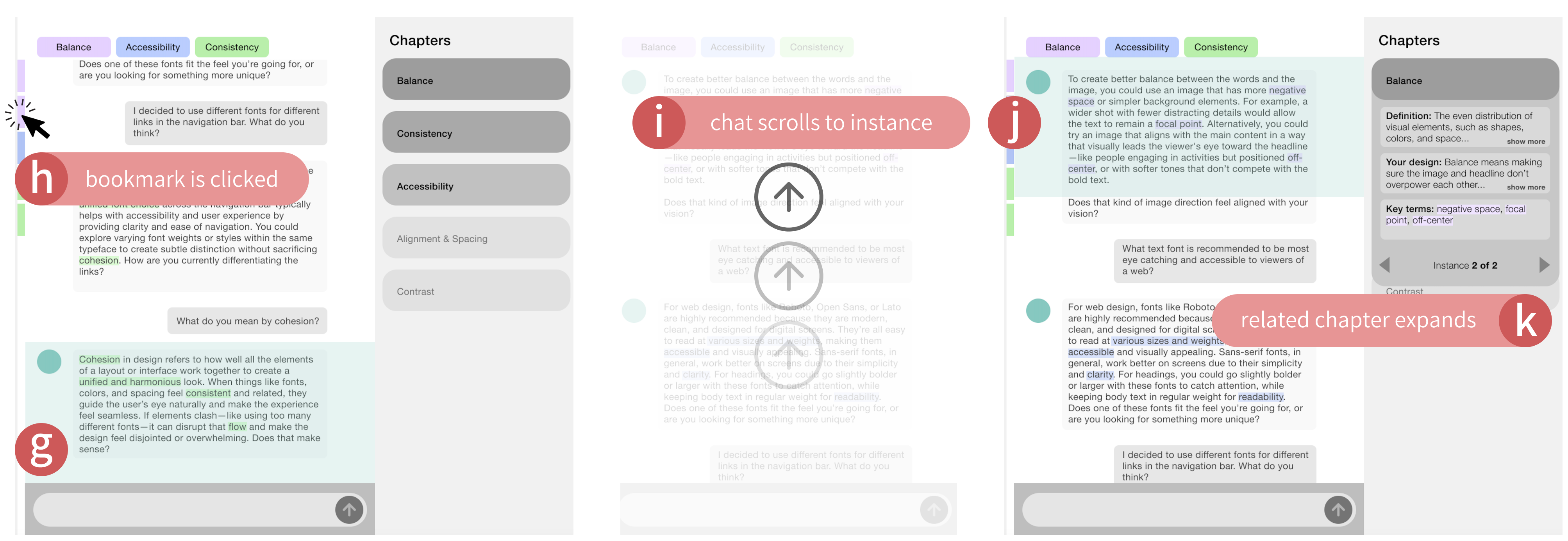}
    \caption{The chat initially anchors at the bottom of the panel \protect\pinkcircled{G} to the most recent message. When the user selects a bookmark \protect\pinkcircled{H}, such as the `Balance' 
    bookmark, the chat automatically scrolls upwards \protect\pinkcircled{I} to the most recent instance where 'Balance' is discussed in the feedback \protect\pinkcircled{J}. At the same time, the corresponding chapter for `Balance' expands in the sidebar \protect\pinkcircled{K}.}
    \label{fig:enter-label}
    \Description{Annotated screenshot demonstrating how selecting a “Balance” bookmark scrolls the chat upward to the relevant feedback. The interface shows the timeline anchored at the latest message (G), and scrolls to the point where the “Balance” principle was last discussed (I and J). At the same time, the corresponding chapter accordion on “Balance” expands in the sidebar (K).}
\end{figure}

\section{User Study}

\subsection{Methodology}

To further develop our vision for layered scaffolding, we conducted a user study with design novices to understand whether and how these features might support the design goals. Given the exploratory nature of this work, we follow a research-through-design method where the focus is on developing design insights as opposed to falsifying hypotheses~\cite{zimmerman2007research}. Participants in the study were asked to think aloud during the study and their responses were analyzed through reflexive thematic analysis~\cite{braun2019reflecting}. The study was approved by IRB.

\textbf{Study Design.}
We designed a fictional conversation using the real user utterances observed in the formative study. We populated the conversation as text in the prototype, enabling users to progress through the design feedback conversation while interacting with the system features.

\textbf{Participants and Recruitment.} We recruited 8 participants (5 female, 2 male, 1 non-binary) from North American colleges and universities. Participants self-identified as beginner designers in the pre-study survey. The exclusion criteria included prior completion of an intermediate-level design course or professional experience as a designer. Participants received a \$20 USD digital gift card upon completing the study.

\textbf{Procedure.} Participants began by completing an informed consent and pre-study survey, followed by analyzing a fictional website mockup and identifying areas for improvement based on visual design principles. They were introduced to five core design principles: \textit{Accessibility, Consistency, Contrast, Balance,} and \textit{Alignment and Spacing}. These principles were explained to help them recognize potential areas for improvement within their designs. After a brief tutorial on navigating the prototype, they engaged with the system while thinking aloud (15 minutes), progressing through the mock conversation and interacting with system features. The session concluded with a post-study survey and semi-structured interview, where participants reflected on their experience.

\subsection{Results}

\subsubsection{Reflections on Learning and Design Principles}

Multiple participants described how their understanding of the design principles improved as a result of using Feedstack. Several participants reported that they began to notice principles they had previously overlooked or not considered relevant. For instance, P6 said:

\begin{quote}
\textit{``Consistency surprised me. It wasn’t something that stood out at first, but now I see how it applies. Also, under balance, things like negative space came up, which I didn’t think of initially.''} 
\end{quote}

Similarly, P4 shared, \textit{``I didn’t notice the balance issue with the image at first. I thought it was fine, but when feedback mentioned it being dominant, it made sense.''} These responses suggest that the system helped surface implicit design principles, encouraging deeper reflection on their applicability.

Participants expressed the value of learning design principles when receiving feedback. For example, P2 shared, 
\begin{quote}
\textit{``Learning about the principles helps change how you approach designing, while specific feedback is more about fixing something. It's like the analogy---if you give someone a fish, they eat for a day, but if you teach them to fish, they can feed themselves. Learning the design principles is just as important as feedback.''}
\end{quote}
Additionally, they shared, \textit{``I think I've overlooked accessibility in design. I usually focus more on aesthetics and less on the practical side of things, so it was interesting to see accessibility as a core consideration.''}

\subsubsection{Desire to Explore Deeper}

Beyond recognizing principles, participants expressed interest in exploring the key terms and highlights. For example, P7 noticed multiple instances of `Balance' throughout the feedback conversation: 

\begin{quote}
\textit{``I just noticed the other instances of balance that are highlighted. It's nice to see how the feedback ties into the principles.''} 
\end{quote}

Similarly, P4 attempted to click on the highlights, expecting a definition. They shared,
\begin{quote}
\textit{``I liked the key terms, though I wished I could click on them to get definitions directly from the paragraph. That's just what I kept trying to do, even though the definitions were already provided [within the feedback].''} 
\end{quote}
This indicates that the user is motivated to engage with the terms at a more granular level, highlighting the potential for the tool to go beyond providing feedback and serve as an interactive learning environment, where students can independently explore and clarify design concepts and connect them to underlying design principles. 

\subsubsection{Continued Use After the Conversation}

Several participants engaged with the system's Chapters after the conversation concluded, engaging with the affordances to reflect on the feedback and further explore the design principles post-conversation. P7 noted, \textit{``It's useful for summarizing, but also for quickly referencing the conversation in more detail. If I want to remember something specific, it's helpful to have the full context behind it.''} This suggests that Feedstack could enable users to revisit feedback at later stages, reinforcing learning and aiding in the iterative design process.

In the context of iteration, participants such as P8 expressed appreciation for the system’s navigational features, sharing that they would use them more after the conversation to implement changes. P8 shared, 
\begin{quote}
\textit{``It’s great for navigating through the conversation without scrolling. I’d probably use this feature more after the conversation when I’m implementing changes.''}
\end{quote}
This indicates that Feedstack’s interactive feedback structure is not only useful for the conversation itself but also supports users in applying feedback and refining their designs. 

\section{Discussion}

\subsection{Synthesizing Design Insights}
\subsubsection{Reflection on Design Principles}

The results of the user study suggest that Feedstack helps novice designers become more aware of design principles and engage with them more deeply. Participants also seemed to be more appreciative of the design principles beyond just the design feedback. These findings are promising and suggest that the provided affordances have the potential to make the implicit aspects of the conversation explicit.

\subsubsection{Support for Exploration and Reflection}

Feedstack appears to encourage active exploration of key terms and design principles. Participants expressed interest in exploring the highlighted key terms more extensively. They also described how Feedstack encouraged them to reflect on the feedback even after the conversation 'ended.' This suggests that the features are helping students go beyond the conversation and engage more holistically with the feedback.

\subsection{Implications for Future Design}

The feedback from participants also points to areas for improvement. For instance, offering more interactive features—such as clickable key terms for direct definitions—could improve users' experience. This could make the tool even more effective in supporting users' learning and promote exploration. 

\begin{figure}
    \centering
    \includegraphics[width=1\linewidth]{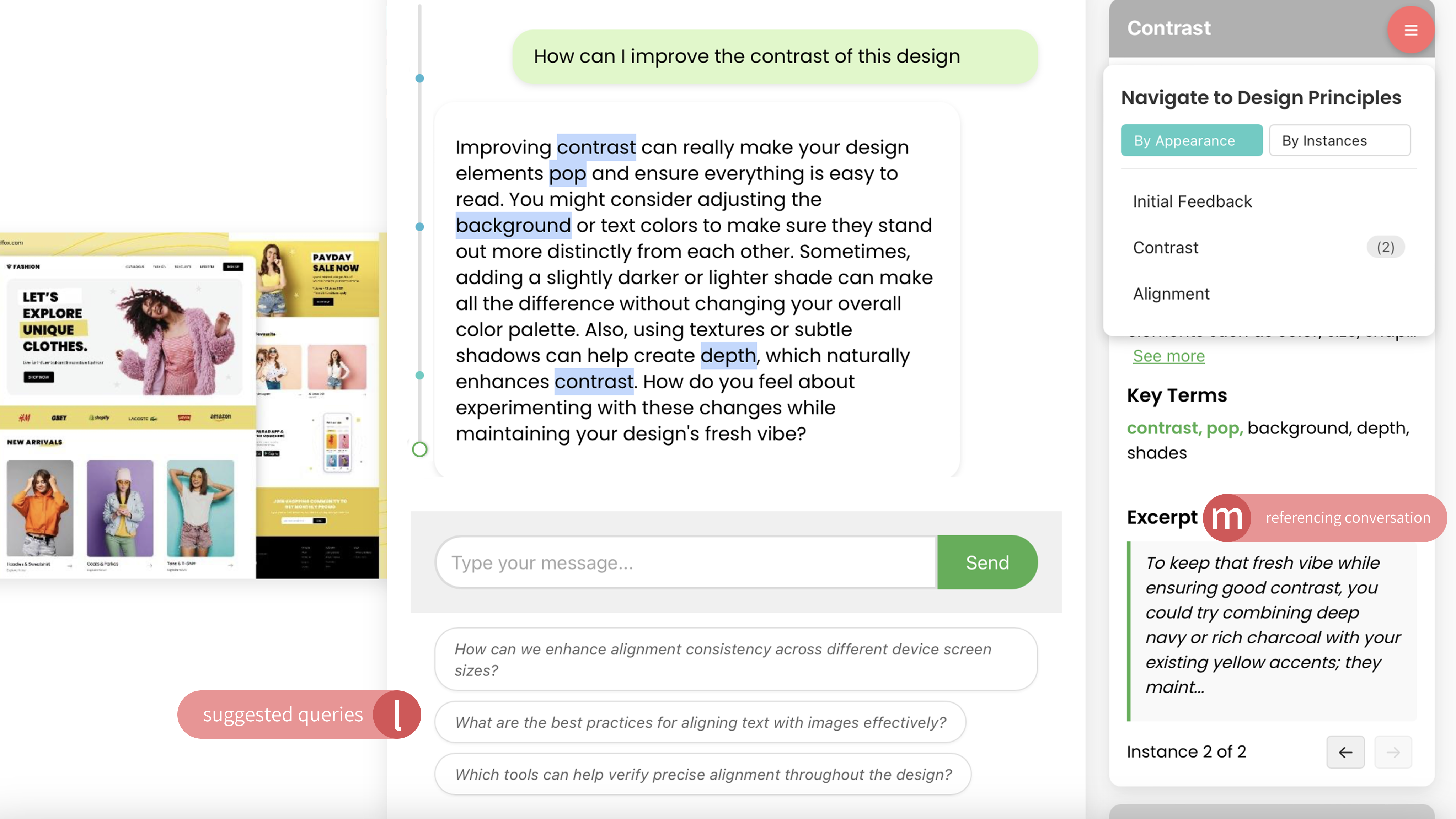}
\caption{The resulting web-based Feedstack system. It incorporates new features informed by recent findings. These include \protect\pinkcircled{L} suggested queries and \protect\pinkcircled{M} excerpt references that highlight the principle being applied during the conversation.}
    \label{fig:system-final}
    \Description{Web-based interface of the updated Feedstack prototype. New features are suggested queries (L) shown near the input area to help guide conversation, and excerpt references (M) which visually indicate where specific principles are being applied or referenced in the chat history.}
\end{figure}

During our user studies, the main issue we observed was challenges in keeping the conversation going. This inspired us to present two additional features  which have been integrated into our functioning prototype. In addition to these features, we also added a feature for referencing feedback instances. The most recent prototype which is implemented in React.js, Django, and uses the OpenAI API is shown in Figure~\ref{fig:system-final}. 

\subsubsection{Emerging Topics} One design idea that emerged from this work was to extend the opacity feature to show suggested topics that are related but have not yet been introduced to the conversation. This provides information scents~\cite{chi2001using} towards concepts and ideas that are just slightly beyond the immediate conversational context. 

\subsubsection{Conversational Cues \protect\pinkcircled{L}} Chatbots can offer conversational cues, such as suggesting conversational turns for the user that they may want to take~\cite{gohsen2023guiding}. This approach could be helpful to guide users toward new topics and ideas. Compared with the \textit{Emerging Topics} feature, we expect this to be more oriented toward feedback instances (clarifying feedback or making it more actionable) as opposed to high-level design principles. 

\subsubsection{Referencing Conversation \protect\pinkcircled{M}} 

Participants explained that they wanted more opportunities to see instances of design feedback for specific design principles. This inspired a navigation area within the Chapters with references to feedback instances in the conversation.

\subsection{Limitations and Future Work} 

Our work is rooted in the research-through-design tradition, where the goal is not to validate hypotheses or to offer conclusive evidence. Rather, the goal is to explore and also to communicate design ideas. The prototype in this study serves as a design probe~\cite{boehner2012probes} to instantiate abstract ideas. However, as future work, the prototype should be evaluated more explicitly with comparisons to existing commercial chatbots. 

\section{Conclusion}

Our research introduces a proof-of-concept interactive system, Feedstack. We explore the use of novel affordances and interactions to structure design feedback conversations, externalizing underlying principles and encouraging users to more deeply engage with them beyond the primary conversation. Our preliminary evaluation indicates that novice designers found value in layered affordances that promote awareness of previously hidden principles. 

\begin{acks}
This research was supported by the National Science Foundation (NSF) under Grant Number 2150152.
\end{acks}
\bibliographystyle{ACM-Reference-Format}
\bibliography{refs}

\end{document}